\input harvmac
\noblackbox

\def\NSNS{${NS}\otimes {NS}$}
\def\RR{${\cal R}\otimes {\cal R}$}

\def\tr{{\rm tr}}

\def\half{{\textstyle {1 \over 2}}}

\def\np{Nucl. Phys.}
\def\pl{Phys. Lett.}

\def\half{{\textstyle {1 \over 2}}}

%%% Paragraphs

%%% special math symbols
 \font\cmss=cmss10
\font\cmsss=cmss10 at 7pt
\def\rlx{\relax\leavevmode}
\def\inbar{\vrule height1.5ex width.4pt depth0pt}
\def\IC{\relax\,\hbox{$\inbar\kern-.3em{\rm C}$}}
\def\IN{\relax{\rm I\kern-.18em N}}
\def\IP{\relax{\rm I\kern-.18em P}}
\def\ZZ{\rlx\leavevmode\ifmmode\mathchoice{\hbox{\cmss Z\kern-.4em Z}}
 {\hbox{\cmss Z\kern-.4em Z}}{\lower.9pt\hbox{\cmsss Z\kern-.36em Z}}
 {\lower1.2pt\hbox{\cmsss Z\kern-.36em Z}}\else{\cmss Z\kern-.4em
 Z}\fi}
%%% misc.
\def\IZ{\relax\ifmmode\mathchoice
{\hbox{\cmss Z\kern-.4em Z}}{\hbox{\cmss Z\kern-.4em Z}}
{\lower.9pt\hbox{\cmsss Z\kern-.4em Z}}
{\lower1.2pt\hbox{\cmsss Z\kern-.4em Z}}\else{\cmss Z\kern-.4em
Z}\fi}

\def\narrowplus{\kern -.04truein + \kern -.03truein}
\def\narrowminus{- \kern -.04truein}
\def\narrowminussub{\kern -.02truein - \kern -.01truein}

\def\half{{1\over 2}}

\def\frac#1#2{{#1\over #2}}

\def\IZ{\relax\ifmmode\mathchoice
{\hbox{\cmss Z\kern-.4em Z}}{\hbox{\cmss Z\kern-.4em Z}}
{\lower.9pt\hbox{\cmsss Z\kern-.4em Z}}
{\lower1.2pt\hbox{\cmsss Z\kern-.4em Z}}\else{\cmss Z\kern-.4em
Z}\fi}
\def\IB{\relax{\rm I\kern-.18em B}}
\def\IC{{\relax\hbox{$\inbar\kern-.3em{\rm C}$}}}
\def\ID{\relax{\rm I\kern-.18em D}}
\def\IE{\relax{\rm I\kern-.18em E}}
\def\IF{\relax{\rm I\kern-.18em F}}
\def\IG{\relax\hbox{$\inbar\kern-.3em{\rm G}$}}
\def\IGa{\relax\hbox{${\rm I}\kern-.18em\Gamma$}}
\def\IH{\relax{\rm I\kern-.18em H}}
\def\II{\relax{\rm I\kern-.18em I}}
\def\IK{\relax{\rm I\kern-.18em K}}
\def\IP{\relax{\rm I\kern-.18em P}}
%\def\IX{\relax{\rm X\kern-.01em X}}
%this doesn't work

\font\cmss=cmss10 \font\cmsss=cmss10 at 7pt
\def\IR{\relax{\rm I\kern-.18em R}}

\def\1{{\bf 1}}
\def\3{{\bf 3}}
\def\7{{\bf 7}}
\def\2{{\bf 2}}
\def\8{{\bf 8}}

%

%
%       \eqn\label{a+b=c}       gives displayed equation, numbered
%                               consecutively within sections.
%     \eqnn and \eqna define labels in advance (of eqalign?)
%
\def\eqnn#1{\xdef #1{(\secsym\the\meqno)}\writedef{#1\leftbracket#1}%
\global\advance\meqno by1\wrlabeL#1}
\def\eqna#1{\xdef #1##1{\hbox{$(\secsym\the\meqno##1)$}}
\writedef{#1\numbersign1\leftbracket#1{\numbersign1}}%
\global\advance\meqno by1\wrlabeL{#1$\{\}$}}
\def\eqn#1#2{\xdef #1{(\secsym\the\meqno)}\writedef{#1\leftbracket#1}%
\global\advance\meqno by1$$#2\eqno#1\eqlabeL#1$$}

\def\tr{{\rm tr}}

\def\half{{\textstyle {1 \over 2}}}

\def\lam16{\lambda^{16}}

\def\hI{\hat I}

\def\lr { \lref}

\parskip=0pt plus 1pt
\parindent 25pt
\tolerance=10000
\lref\bvs{M. Bershadsky, C. Vafa and V. Sadov, 
{\it D-branes and topological field theories}, Nucl. Phys. 
{\bf B463} (1996) 420; hep-th/9511222.}
\lref\fadd{L. Faddeev, {\it Operator anomaly for the Gauss law}, 
Phys. Lett. {\bf 145B} (1984) 81.}
\lref\dscnti{M.F. Atiyah and I.M. Singer,
{\it Dirac operators coupled to vector bundles},
Proc. Natl. Acad. Sci. {\bf 81} (1984) 2597.}
\lref\dscntii{
L. Faddeev and S. Shatashvili, {\it Algebraic and hamiltonian methods
in the theory of nonabelian anomalies}, Theor. Math. Fiz., {\bf 60}
(1984) 206; english transl. Theor. Math. Phys. {\bf 60} (1984) 770.}
\lref\dscntiii{B. Zumino, {\it Chiral anomalies and differential
geometry}, in ``Relativity, Groups and Topology II,'' proceedings of
the Les Houches summer school, Bryce S DeWitt and Raymond Stora, eds. 
North-Holland, 1984.}
\lref\dscntiv{
For reviews see ``Symposium on Anomalies, Geometry and Topology,''
William A. Bardeen and Alan R. White, eds. World Scientific, 1985, and
L. Alvarez-Gaum\'e and P. Ginsparg,
{\it The structure of gauge and gravitational anomalies},
Ann. Phys. {\bf 161} (1985) 423.}
\lr\greenb{M.B.  Green and J.H. Schwarz,  
{\it Extended supergravity in ten dimensions},
Phys. Lett. {\bf 122B} (1983) 143.}
\lref\schwarzwest{J.H.  Schwarz and P.C.  West, {\it Symmetries and
transformations of chiral $N=2$ $D=10$ supergravity}, Phys. Lett. {\bf
126B} (1983) 301.}
\lref\greennem{ M.B. Green and D. Nemeschansky (unpublished); 
Presented by D. Nemeschansky  at PASCHOS'95, Baltimore, April 1995
and by M.B. Green at SUSY'95, Paris, June 1995.}
\lr\greena{
M.B.  Green and  J.H.  Schwarz,
{\it Covariant description of superstrings}, 
Phys.  Lett. {\bf 136B} (1984) 367.}
\lr\greenschwarza{
M.B.  Green and  J.H.  Schwarz,
{\it Anomaly cancellations in supersymmetric $D=10$ gauge theory and
superstring theory}, Phys. Lett. {\bf 149B} (1984) 117.}
\lr\greene{
B.R. Greene, A. Shapere, C. Vafa and S.T. Yau, 
{\it Stringy cosmic strings and noncompact Calabi-Yau manifolds},
Nucl. Phys. {\bf B337} (1990) 1.}
\lr\schwarzb{J.H. Schwarz, 
{\it Covariant field equations of chiral $N=10$, $D=10$ supergravity},
\np\ {\bf B226} (1983) 269.}
\lr\howea{P.S. Howe and P.C. West, 
{\it The complete $N=2$, $d=10$ supergravity},
\np\ {\bf B238} (1984) 181.}
\lr\hulla{C.M.  Hull and P.K.  Townsend, {\it The unity of string
dualities}, \np\ {\bf B438} (1995) 109; hep-th/9410167.}
\lr\grisarua{M.T. Grisaru, P. Howe, L. Mezincescu, B.E.W.  Nilsson
and P.K.Townsend, 
{\it $N=2$ Superstrings in a supergravity background}, 
\pl\ {\bf 162B} (1985) 116.}
\lr\alvarezga{L. Alvarez--Gaum\'e and E.  Witten,
{\it Gravitational Anomalies }, 
Nucl. Phys. {\bf B234} (1984) 269.}
\lr\marcusb{N.  Marcus,
{\it Composite anomalies in supergravity},
Phys.  Lett. {\bf  157B} (1985) 383.}
\lr\feynmang{R.  Endo and M.  Takao, 
{\it Feynman gauge for the Rarita--Schwinger field in higher
dimensions and chiral $U(1)$ anomaly}, 
Phys.  Lett. {\bf 161B} (1985) 155.}
\lr\popea{C.J. Isham and C.N. Pope, 
{\it Nowhere-vanishing spinors and topological obstructions to the
equivalence of the NSR and GS superstrings}, 
Class. Quant. Grav. {\bf 5}  (1985) 257.}  
\lr\calharv{C.G.  Callan and J.A.  Harvey {\it Anomalies and fermion
zero modes on strings and domain walls}, 
Nucl. Phys. {\bf B250} (1985) 427.}
\lr\luduff{M.J. Duff, J.T. Liu and  R. Minasian, 
{\it Eleven-dimensional origin of string-string duality: A One loop
test}, Nucl. Phys. {\bf B452} (1995) 261; hep-th/9506126.}
\lr\vafwitta{C. Vafa and E.  Witten, 
{\it A one-loop test of string-string duality}, 
Nucl. Phys. {\bf B447} (1995) 261; hep-th/9505053.}
\lr\sensb{A. Sen, {\it  Duality and orbifolds}, 
Nucl.Phys. {\bf B474} (1996) 361; hep-th/9604070.}
\lr\beckera{K. Becker and M. Becker, 
{\it M-Theory on Eight-Manifolds},
Nucl.Phys. {\bf B477} (1996) 155; hep-th/9605053.}
\lr\vafwittb{S. Sethi, C. Vafa and E. Witten, 
{\it Constraints on Low-Dimensional String Compactifications},
Nucl.Phys. {\bf B480} (1996) 213; hep-th/9606122.}
\lr\sena{A.~Sen, 
{\it Strong-Weak coupling duality in four dimensional string theory},
Int J. Mod. Phys. {\bf A9} (1994) 3707.}
\lr\erdelyi{A. Erdelyi {\it et al}, 
{\it Higher transcendental functions},
McGraw-Hill, New York, 1953.}
\lr\greenc{M.B. Green, J. Harvey and G. Moore, 
{\it I-Brane inflow and anomalous couplings on D-branes},
Class. Quant. Grav. {\bf 14} (1997) 47; hep-th/9605033.}
\lr\dasguptaa{K. Dasgupta, D.P. Jatkar and S. Mukhi,
{\it Gravitational couplings and $Z_2$ orientifolds}, 
Nucl. Phys. {\bf B523} (1998) 465; hep-th/9707224.}
\lr\dasguptab{K. Dasgupta and S. Mukhi,
{\it Anomaly Inflow on orientifold planes},
JHEP {\bf 03} (1998) 004; hep-th/9709219.}
\lr\vafa{C.  Vafa, {\it Evidence for F theory},  Nucl. Phys. 
{\bf B469} (1996) 403;  hep-th/9602022.} 
\lr\senb{A. Sen,
{\it Strong coupling dynamics of branes from M-theory}, 
JHEP {\bf 10} (1997) 002; hep-th/9708002.}
\lr\milnor{J.W. Milnor and J.D. Stasheff, ``Characteristic
classes,'' Annales of Mathematics Studies 76, Princeton University 
Press (1974).}
\lr\joyce{D.D. Joyce, {\it Compact 8-manifolds with holonomy
$Spin(7)$}, Invent. Math. {\bf 123} (1996) 507.}
\lr\shatashvili{M. Lippert, B. Freivogel and S.  Shatashvili, 
{\it Manifolds of exceptional holonomy and string dualities}, Yale
preprint (in preparation).} 
\lr\salama{A. Salam and E.  Sezgin, {\it $d=8$ supergravity: matter
couplings, gauging and Minkowski 
compactification}, Phys.  Lett. {\bf 154B} (1985) 37.}
\lr\townsends{M. Awada and P.K.  Townsend, {\it $d=8$
Maxwell--Einstein supergravity}, Phys.  Lett. {\bf 156B} (1985) 51.}
\lr\sendd{A.  Sen, {\it A non-perturbative description of the
Gimon--Polchinski orientifold},  hep-th/9611186;
 Nucl. Phys. {\bf B489} (1997) 139.}
\lr\senee{A.  Sen, {\it F theory and the Gimon--Polchinski
orientifold}, hep-th/9702061;  Nucl. Phys. {\bf B498} (1997) 135.}

\Title{\vbox{\baselineskip8pt
\hbox{hep-th/9810153}
\hbox{DAMTP-98-102}
\hbox{}
}}
{\vbox{
\centerline{An $SL(2,{{\rm Z}})$ anomaly in IIB
supergravity }
\vskip 0.2cm
\centerline{and its F-theory interpretation}}}
\bigskip
 \centerline{ Matthias R. Gaberdiel\foot{email:
M.R.Gaberdiel@damtp.cam.ac.uk}}
\smallskip
\centerline{and} 
\smallskip
\centerline{Michael B.  Green\foot{email: M.B.Green@damtp.cam.ac.uk}}
\vskip 0.05in
\centerline{\it Department of Applied Mathematics and
Theoretical Physics}
\centerline{\it Silver Street,  Cambridge CB3 9EW, UK}

\vskip 1.0cm

The $SL(2,\IZ)$ duality transformations of type IIB supergravity are
shown to be anomalous in generic F-theory backgrounds due to the
anomalous transformation of the phase of the chiral fermion
determinant. The anomaly is partially cancelled provided the
ten-dimensional type IIB theory lagrangian contains a term that is a
ten-form made out of the composite $U(1)$ field strength and four
powers of the curvature. A residual anomaly remains uncancelled, and
this implies a certain topological restriction on consistent
backgrounds of the euclidean theory. A similar, but slightly stronger,
restriction is also derived from an explicit F-theory compactification
on $K_3 \times M^8$ (where $M^8$ is an eight-manifold with a nowhere
vanishing chiral spinor) where the cancellation  of tadpoles for
Ramond--Ramond fields is only possible if $M^8$ has an  Euler
character that is  a positive multiple of $24$. The interpretation of
this restriction in the dual heterotic theory on $T^2 \times M^8$ is
also given.

%\draftmode
\vskip 0.1in
\Date{8/98 (revised 2/99)}

\newsec{Introduction}
The unravelling of the discrete local gauge symmetries of string
theory may provide significant clues concerning its underlying
geometric structure. We will here be concerned with how the $SL(2,\IZ)$
symmetry of type IIB string theory is consistently realized upon 
compactification.  

Classical type IIB supergravity
\refs{\greenb,\schwarzwest,\schwarzb,\howea} is invariant under a
global $SL(2,\IR)$ symmetry. In a gauge invariant formulation of the
theory there are three scalar fields that parametrise an $SL(2,\IR)$
matrix while the chiral fermionic fields transform under a local
$U(1)$ (or $O(2)$) symmetry but are inert under the global $SL(2,\IR)$
transformations.  Upon fixing the gauge the $U(1)$ is identified with
a subgroup of $SL(2,\IR)$ and one of the scalars is eliminated.  The
two remaining scalars parametrise the coset space $SL(2,\IR)/U(1)$, or
the upper-half plane.  In this case a general $SL(2,\IR)$
transformation must be accompanied by a compensating $U(1)$ gauge
transformation in order to maintain the gauge condition. This induces
a nontrivial $SL(2,\IR)$ transformation on the fermion fields.  Of
course, the consistency of this procedure requires that the $U(1)$
symmetry is not anomalous.  The structure of the IIB theory and the
gauge fixing procedure will be reviewed in section 2.

The continuous $SL(2,\IR)$ symmetry of the classical supergravity does
not survive in the quantum theory since it is not preserved in the
string extension of type IIB supergravity.  Indeed it is well known
\refs{\greena} that the classical superstring is not invariant under
the $U(1)$ subgroup of $SL(2,\IR)$ that rotates the two supercharges
into each other.  Since the two supercharges move in opposite
directions on the world-sheet the theory is only invariant under a
discrete $\IZ_4$ subgroup of this $U(1)$ which interchanges the two
supercharges {\it and} reverses the direction $\sigma$. It is also
well-established by now that the $SL(2,\IR)$ symmetry of classical IIB
supergravity is actually replaced by a discrete $SL(2,\IZ)$ 
{\it local} symmetry in string theory \refs{\hulla}.  Thus, the moduli
space of the scalar fields becomes 
$SL(2,\IZ)\backslash SL(2,\IR)/U(1)$ and the $\IZ_4$ symmetry that
acts on the supercharges is the intersection of $U(1)$ with
$SL(2,\IZ)$.  

In this paper we analyse an anomaly in the action of the $SL(2,\IZ)$
duality symmetry group. This anomaly can be understood to originate
from an anomaly in the conservation of the local chiral $U(1)$ current
that will be exhibited in section 3.\foot{ This $U(1)$
anomaly was originally obtained in \refs{\greennem}.}
There we will evaluate an anomalous hexagon diagram coupling the
divergence of the current to four gravitons and to the current
itself. As expected, the anomaly is cancelled provided the action
contains a certain local interaction, $S'$. This term breaks
the $SL(2,\IR)$ symmetry of the theory, but the anomaly of 
an $SL(2,\IZ)$ subgroup is partially cancelled provided that the
ten-dimensional type IIB action contains a term of the form 
\eqn\ctt{S'' = {i \over 4\pi}
\int d^{10}x\, \ln g(\tau,\bar\tau) \, F \wedge X_8(R)\,.}
Here $F$ is the $U(1)$- field strength, $X_8(R)$ is an eight-form
linear combination of Pontryagin classes, and $g$ is a function of
$\tau$ that is not determined uniquely by these considerations alone.
This term cancels the $SL(2,\IZ)$ anomaly up to $\tau$-independent
terms, and the condition that the whole $SL(2,\IZ)$ symmetry is
unbroken becomes 
\eqn\condd{ {1\over 4\pi} \int d^{10}x \, F \wedge X_8(R) \in 4 \IZ\,.}
For the case of an F-theory compactification on the product of an
elliptically fibered $K_3$ tensored with a euclidean eight-manifold,
$M^8$, \condd\ simplifies to the condition that the Euler character of
$M^8$ is a multiple of $24$. This condition is reminiscent of the
conditions for consistent compactifications of type IIA superstring
theory to two dimensions \refs{\sensb}, M-theory to three dimensions
\refs{\beckera} and F-theory to four dimensions \refs{\vafwittb},
where the consistent eight-manifolds were restricted to have Euler
numbers that are positive multiples of $24$ (although our anomaly
considerations do not, by themselves, require positivity of the Euler
number).  
 
These F-theory compactifications raise further consistency
questions concerning the cancellation of tadpoles that arise from
wrapping D-branes around homology cycles. We will study this in some
detail in section 4 for the example of F-theory compactified on
$K_3\times M^8$.  In that case there are $24$ seven-branes that wrap
around $M^8$ and give tadpoles for $C^{(4)}$ and $C^{(0)}$ (the
Ramond--Ramond, or \RR, four-form and zero-form potentials).  These
tadpoles must be cancelled by adding a (integer) number of
three-branes and D-instantons.  The condition that this is possible is
once again that $\chi$ be a multiple of $24$. However, now $\chi$ is
also required to be positive and, in fact, the number of D-instantons
is precisely $\chi/24$. 

Both the anomaly and the tadpole analysis also have analogues in terms
of the heterotic string compactified on $T^2$ which will be discussed
in section 5.

\newsec{ Fields and gauge fixing of type IIB
supergravity}

The covariant field equations of IIB supergravity are invariant under
global $SL(2,\IR)$ transformations \refs{\schwarzb,\howea}. There is a
well-known problem in formulating a globally well-defined lagrangian
for the IIB theory due to the presence of the self-dual five-form
field strength. However, since this field does not play a r\^ole in
the following, we can proceed as if there were a lagrangian (although
the description of the anomaly does not actually 
require an explicit lagrangian
\refs{\alvarezga}).  

\subsec{Symmetries and fields}

The scalar field coset space $SL(2,\IR)/U(1)$ may be described in the
usual manner by a zweibein, $V^\alpha_i$ ($\alpha, i=1,2$), with
components,
\eqn\beindef{{1\over \sqrt{\tau_2}} \pmatrix{\tau_2 \cos \phi + \tau_1
\sin\phi & -\tau_2 \sin\phi +\tau_1 \cos\phi \cr 
\sin \phi & \cos\phi  },}
which is a $SL(2,\IR)$
matrix. Introducing a complex basis, 
\eqn\noname{
V_{\pm}^\alpha = (-2 i )^{-1/2} 
\left(V_2^\alpha \pm i V_1^\alpha \right) \,,}
$V$ may be written as 
\eqn\vinlamdef{V^\alpha_\pm  = \pmatrix{V^1_- & V^1_+ \cr
V^2_- & V^2_+\cr}=(-2i\tau_2)^{-1/2}\pmatrix{
\bar\tau e^{-i\phi} & \tau e^{i\phi} \cr
e^{-i\phi} & e^{i\phi}  & \cr}\,. }
The complex scalar $\tau$ parametrises the Poincar\'e upper-half
plane and $\phi$ is an angular field ($0\le \phi\le 2\pi$).  The
global $SL(2,\IR)$ transformations act on the matrix $V$ from the left
(so that $V_\pm^\alpha$ transforms as a vector). The $U(1)$ subgroup
of $SL(2,\IR)$ that acts locally consists of the real matrices 
\eqn\otwodef{ \pmatrix{ \cos{\Sigma} & - \sin{\Sigma} \cr
\sin{\Sigma} & \cos{\Sigma}\cr}\,,}
which are rewritten in the complex basis as
\eqn\uonedef{ \pmatrix{e^{-i\Sigma} & 0 \cr 0 &
e^{i\Sigma}\cr}\,,} 
where $0\leq \Sigma \leq 2 \pi$ is an angle; this subgroup acts on
$V$ from the right (and thus on the index $\pm$ as 
$V_{\pm}^{\alpha} \rightarrow V_{\pm}^{\alpha} e^{\pm i \Sigma}$).
Therefore, a 
general $SL(2,\IR) \times U(1)$ transformation acting on $V$ gives 
\eqn\gentrans{V^{\prime \alpha}_\pm =
U^\alpha_{\ \beta} V^\beta_\pm e^{\pm i
\Sigma(x)}\,,}  
where $U^\alpha_{\ \beta}$ is a constant $SL(2,\IR)$ matrix,
\eqn\suonemat{\pmatrix{ a & b \cr
 c & d \cr}\,, }
and $a$, $b$, $c$ and $d$ are real numbers satisfying
$ad-bc =1$. Gauge invariance of the classical  
theory will allow one of the three scalars to be eliminated from
$V^\alpha_\pm$.

The $SL(2,\IR)$-singlet combination,
\eqn\gaugepot{\eqalign{Q_\mu =& -i\epsilon_{\alpha\beta} 
V_-^\alpha\partial_\mu  V_+^\beta
\cr =& \partial_\mu \phi - {\partial_\mu \tau_1\over
2\tau_2}\,,\cr}}
acts as a composite gauge potential since it transforms as 
$\delta Q_\mu = \partial_\mu \Sigma$ under the local $U(1)$
transformation in \gentrans. The scalar fields also package together
into a complex $SL(2,\IR)$-invariant combination, $P^\mu$, defined by
\eqn\pdef{\eqalign{P_\mu = & -\epsilon_{\alpha\beta} V_+^\alpha 
\partial_\mu V_+^\beta \cr  = 
& i e^{2i\phi} {\partial_\mu \tau \over 2\tau_2} 
\,,\cr}}
which transforms under $U(1)$ as $P_\mu \to  e^{2i\Sigma} P_\mu$.
This  means that its $U(1)$ charge is 2, while the complex conjugate
field, $P^*$, has charge $-2$. The abelian field strength satisfies
\eqn\pqrel{F\equiv dQ = -i P\wedge P^* 
= {id \bar\tau \wedge  d\tau \over 4 \tau_2^2}\,,} 
which follows from the definitions \gaugepot\ and \pdef.

The other bosonic fields of type IIB supergravity are neutral under
the local $U(1)$. The two components of the second-rank antisymmetric
tensor potential, $A^\alpha_{\mu\nu}$, form a $SL(2,\IR)$ vector 
and transform as 
$A^{\prime \alpha}_{\mu\nu} = U^\alpha_{\ \beta} A^\beta_{\mu\nu}$.
The real  metric, $g_{\mu\nu}$, is neutral under both the $U(1)$ and
the $SL(2,\IR)$ as is the real fourth-rank antisymmetric tensor
potential, $C^{(4)}$.

Turning to the fermions, the two gravitini form a complex conjugate
pair of spin-3/2 gravitino fields, $\psi_\mu$ and $\psi_\mu^*$, with
$U(1)$ charges  
$\pm 1/2$, while the two spin-1/2 dilatino 
fields form a complex conjugate pair
of opposite chirality, $\lambda$ and $\lambda^*$, with $U(1)$ charges 
$\pm 3/2$, 
\eqn\spinortrans{\delta \psi_\mu = {1\over 2} i \Sigma
\psi_\mu,\qquad\qquad\delta \lambda = - {3\over 2} i \Sigma \lambda\,.} 

\subsec{$U(1)$ gauge fixing}

In the classical supergravity theory the local $U(1)$ symmetry can be
used to eliminate the field $\phi$, one of the three scalar fields
that parametrise the $SL(2,\IR)$ group manifold. This is achieved by a
choice of `gauge fixing condition' $\phi =  \hat\phi(\tau)$. In the
gauge fixed theory, the field $V^\alpha_\pm$ transforms under a global
$SL(2,\IR)$ transformation as in \gentrans, where $\Sigma(x)$ is
chosen so as to reinstate the gauge fixing condition. More explicitly,
under the action of $A$ we find that 
\eqn\aact{
A V^\alpha_\pm = ( - 2 i (A\tau)_2)^{-1/2} 
\pmatrix{ (A \tau)^\ast e^{i(\gamma - \hat\phi(\tau))} & 
A \tau e^{i(\hat\phi(\tau) - \gamma)} \cr 
e^{i(\gamma - \hat\phi(\tau))} & 
e^{i(\hat\phi(\tau) - \gamma)} \cr} \,,}
where $A\tau$ denotes the standard action of $SL(2,\IR)$ on the
modular parameter $\tau$,
\eqn\modtrans{ A \tau = {a \tau + b \over c \tau + d} \qquad
\hbox{where} \quad A= \pmatrix{a & b \cr c & d}\,,}
and 
\eqn\gamfed{
e^{i\gamma(A,\tau)} = \left({c\bar\tau +d\over c\tau+d}
\right)^{1/2}\,.} 
The compensating gauge transformation is therefore 
\eqn\compensate{
e^{i\Sigma_A} = e^{i ( \gamma(A,\tau) + \hat\phi(A\tau) 
- \hat\phi(\tau))} \,, }
and the scalars transform as in \modtrans. In general, the
compensating $U(1)$ transformation \compensate\ is non-trivial, and
since the local $U(1)$ acts on the fermions, they themselves transform
non-trivially under $SL(2,\IR)$ transformations.
\vskip4pt

The above construction is covariant under the adjoint action of
$SL(2,\IR)$ since  if we gauge by an $U(1)$ subgroup that is obtained
from \otwodef\ by conjugation by an element $M\in SL(2,\IR)$, 
$S'(\Sigma) = M^{-1} S(\Sigma) M$, then the action of 
$A\in SL(2,\IR)$ on $\tau'$ (where $\tau'$ is defined by 
$M V(\tau,\hat\phi) M^{-1}=V(\tau',\hat\phi')$) is
given as  
\eqn\primedef{
\tau' \mapsto {a' \tau' + b' \over c' \tau' + d'} \qquad
\hbox{where} \qquad 
M A M^{-1} = \pmatrix{a' & b' \cr c' & d' \cr} \,.
}

\newsec{The anomaly}

A striking feature of type IIB supergravity (and superstring theory)
is the fact that all the local gravitational anomalies cancel
\refs{\alvarezga}. However, the question of anomalies in  the (chiral)
$U(1)$ symmetry requires separate attention. A space-time dependent 
$U(1)$ anomaly would be a disaster since it would  prevent the
elimination of the redundant bosonic field and hence  would change
the particle content of the theory. One way to see that this 
cannot happen is to use the method of descent
\refs{\fadd,\dscnti,\dscntii,\dscntiii,\dscntiv} from twelve
dimensions together with the 
fact that there is no appropriate invariant twelve-form
\refs{\marcusb}. The obvious candidate  twelve-forms would have been
$\tr R^4 \wedge F^2$ and $(\tr R^2)^2\wedge F^2$, where the curvature
two-form is defined in the usual manner. However, due to the fact that
$F = -iP \wedge P^*$, these expressions vanish identically.  The
analysis of the anomaly below  will show directly how the potential
local anomaly is cancelled by a local counterterm.

To begin with,  consider the variation of the generating function under
a local $U(1)$ transformation, which may lead to a non-zero
divergence of the $U(1)$ current,  
\eqn\curranom{J^\mu = {1\over 2} \bar \psi_\nu \gamma^{\nu\mu\rho}
\psi_\rho  + {3\over 2}  \bar \lambda \gamma^\mu \lambda\,.}
At the linearized level the one-loop diagrams that can give anomalous
results must have at least six vertices. The particles circulating
around the diagrams are the chiral fermion fields, $\psi_\mu$ and
$\lambda$. Since the chiral ten-dimensional self-dual antisymmetric
tensor is neutral under the $U(1)$ it does not contribute to the
anomaly. In a general gauge there may be contributions from diagrams
with four external gravitons one $P$ line and one $P^*$ line in
addition to the vertex coupling to $\partial_\mu J^\mu$; this is a
{\it heptagon} diagram. Since there is a vertex in the theory that
couples a $P$ to a $\psi$ and $\lambda$ field such diagrams are rather
complicated. There are also familiar hexagon diagrams with
$\partial_\mu J^\mu$ coupling to four external gravitons and one
external $Q$ vertex, which couples to the $U(1)$ charge of the chiral
fermions. Apart from the fact that $Q_\mu$ is a composite gauge field
the structure of these diagrams is very similar to the mixed anomaly
diagrams considered in \refs{\alvarezga}. The complete anomaly is
given by  the sum of diagrams with all permutations of the external
legs.  

Things simplify in the Feynman gauge for the gravitino (which is the
gauge implicitly used in \refs{\alvarezga} and formulated more
explicitly in \refs{\feynmang}). This is the gauge in which the
gravitino propagator is $\Delta_{\mu\nu} = \delta_{\mu\nu}
(\gamma\cdot p)^{-1}$ and in which there is a bosonic spin-$\half$
ghost field that has the opposite chirality but the same $U(1)$ charge
as the gravitino. In this gauge the heptagon diagrams do not
contribute to the anomaly and the only contributions come from the
expected hexagon diagrams.

The calculation is standard and leads to an anomalous phase of the
variation of the partition function under a local $U(1)$
transformation which is given by  
\eqn\totanom{\Delta  =  -{ 1\over (2\pi)^5} \int d^{10} x 
\left({3\over 2} \hI_{1/2}\left(R, 3F/2 \right) - {1\over 2} 
\hI^{(D=10)}_{3/2}\left(R,F/2\right)\right)\Sigma(x)\,,}
where the coefficients arise from the $U(1)$ charges that enter at two
vertices in the hexagon diagram and $\hI$ is defined as in
\refs{\alvarezga}. The integral is a linear combination of the 
index of the twisted Dirac operator acting on a charge-$3/2$  Dirac
fermion (with density $\hI_{1/2}(R,3F/2)$) and on  a charge-$1/2$
Rarita--Schwinger fermion (with density $\hI^{(D=10)}_{3/2}(R,F/2)$),
which is defined to include the contribution of the Fadeev--Popov
ghost and where the superscript indicates the dimension of the tangent
space). Since the  integration is ten-dimensional and since
$F^n\equiv 0$ for $n>1$, only ten-forms linear in $F$ in the expansion
contribute, 
\eqn\diracgen{\eqalign{\hI_{1/2}(R, F) \equiv &
\hI_{1/2}(R)\wedge F \cr   
= &{1\over 5760}(7p_1^2 - 4p_2)\wedge F\,,\cr}} 
where the $p_n$
are the usual Pontryagin polynomials, 
\eqn\pdef{p_1 = \sum x_i^2 = -{1\over 2}\tr R^2, \qquad  
p_2 = \sum_{i<j} x_i^2x_j^2 
= -{1\over 4}\tr R^4 + {1\over 8} (\tr R^2)^2}
and $x_i$ are the skew eigenvalues of the curvature two form, which is
a $SO(10)$ matrix. The expression $\hI_{1/2}(R)$ is the eight-form
contribution to the Dirac genus. The contribution from the chiral
gravitino field combined with its bosonic ghost (of the same $U(1)$
charge) is given by \refs{\alvarezga}  (taking care to set $D=10$ in
the dimension-dependent term) 
\eqn\gravigen{\eqalign{\hI^{(D=10)}_{3/2}(R,F) \equiv 
&  \hI^{(D=10)}_{3/2}(R) \wedge F \cr
 = & {1\over 5760} (303 p_1^2 - 996 p_2) \wedge F\,.
\cr}}
Substituting these expressions into \totanom\ and reexpressing the
$p_i$ in terms of the curvatures gives the local $U(1)$ anomaly in the
form  
\eqn\totanom{\eqalign{\Delta & = -{1\over (2\pi)^5}\int d^{10}x
\left({9\over 4} \hI_{1/2} - {1\over 4} \hI^{(D=8)}_{3/2}\right)
\wedge F\, \Sigma(x) \cr
& = 
{1\over (2\pi)^5} {1\over 96} \int (4p_2 - p_1^2)\wedge F\,
\Sigma(x)\cr
&  =- { 1\over (2\pi)^5}{1\over 96} \int d^{10} x \left(\tr
R^4 - {1\over 4} (\tr R^2)^2\right)\wedge  F\, \Sigma(x)
\,.\cr}}

It is quite striking that the eight-form part of the anomaly in
\totanom\  is proportional to the eight-form that arises as the
coefficient of a $B_{\mu\nu}$ tadpole in the type IIA theory
\refs{\vafwitta}. Superficially, the origin of the eight-form in
\totanom\  in the present calculation is entirely different from its
origin in the IIA theory where  it can be obtained, 
via the anomaly inflow argument \refs{\calharv}, from the cancellation of
a chiral gravitational anomaly in the five-brane of the type IIA
theory \refs{\luduff}. That local anomaly arises from a sum of
terms due to the 
presence of two six-dimensional chiral fermions and one
six-dimensional self-dual antisymmetric tensor potential.  The anomaly
due to the two fermions descends from the eight-form
$2\hI_{1/2}$ while that of the self-dual antisymmetric tensor
descends from the eight-form $\hI_A$ that is related to the 
signature of the manifold. The total anomaly in the five-brane
therefore descends from the expression 
\eqn\expvaf{-{1\over (2\pi)^4} (2\hI_{1/2} + \hI_A)
\,.}
It is possible to eliminate  $\hI_A$ in favour of the index of the
spin-3/2 operator by using the identity 
$21\hI_{1/2} - \hI^{(D=6)}_{3/2} + 8 \hI_A=0$ which was noted in
the discussion of gravitational anomalies in  \refs{\alvarezga},  for
the case in which  the dimension-dependent term in $\hI_{3/2}$ is
set to $D=6$ (the same  dimensionality as the five-brane world
volume).  However, in our context the index  $\hI^{(D=10)}_{3/2}$ in
\gravigen\ is defined with  $D=10$ and the identity  becomes
$25 \hI_{1/2} - \hI^{(D=10)}_{3/2} + 8\hI_A =0$.  
Using this to eliminate $\hI_A$ from \expvaf\ gives the expression 
\eqn\vafwitex{{1\over (2\pi)^4 \cdot  8}\left(9
\hI_{1/2} - \hI^{(D=10)}_{3/2}\right)\equiv -{X_8 \over 4}\, ,} 
that enters as a factor in \totanom.

The above anomaly calculation is based on the weak coupling IIB
Feynman rules, but since the value of the anomaly does not depend on
the coupling constant it presumably has validity beyond perturbation
theory. In fact, since the value of the anomaly depends on $F$ which
is a function of the scalar field $\tau$ that vanishes for constant
$\tau$, non-zero values for the anomaly generically arise from
configurations in which $\tau$ cannot be restricted to small coupling.

\subsec{Local $U(1)$ anomaly cancellation}

As expected, the local $U(1)$ anomaly can be cancelled by adding a local
term to the action,
\eqn\anomcancel{S' = {1\over (4\pi)} \int d^{10} x \,
\phi \, F \wedge X_8(R)\,,}
where $X_8$ is the eight-form defined by \vafwitex, 
\eqn\xeightd{X_8 = {1\over (2\pi)^4} {1\over 48} \left({\rm tr}\,  R^4
- {1\over 4} ({\rm tr} \, R^2)^2\right)}
and in our conventions the partition function is defined by the
functional integral over the fields $\Phi$, 
\eqn\partdef{
Z = \int D\Phi\, e^{i S[\Phi]}\,.}
There is an ambiguity in the definition of \anomcancel\ since $\phi$
is a periodic variable of period $2\pi$.  The consistent
interpretation of the functional integral in the presence of this
term therefore requires  
\eqn\consist{ {1\over 4\pi} \int d^{10} x \,
F \wedge X_8(R) \in \IZ\,,}
a condition which, as we  will see shortly,  
is automatically satisfied by consistent backgrounds.

Under a local $U(1)$ transformation
\eqn\locph{\phi(x) \to \phi(x) + \Sigma (x),}
while the other fields in $S'$ are inert. Therefore 
$\delta S' = - \Delta$ and the local anomaly is cancelled as
required. However, the term $S'$ is not $SL(2,\IR)$ invariant
since  it follows from \aact\ and \gamfed\ that the field $\phi$
transforms as 
\eqn\phitrans{\delta \phi = - {i\over 2} 
\ln\left({c\tau + d\over c\bar\tau + d}\right),}
under $SL(2,\IR)$ while the other fields in $S'$ are inert. The
corresponding anomaly as well as the expression in \consist\ vanish
automatically provided that the scalar fields are constant throughout
space since then $F=0$. The interesting configurations thus only arise
for $F\ne 0$ which is characteristic of F-theory compactifications.

\subsec{$SL(2,\IZ)$ anomaly cancellation}

A priori one should not expect that the quantum theory maintains the
whole $SL(2,\IR)$ symmetry, but only that an $SL(2,\IZ)$ subgroup is
unbroken. The anomaly in this subgroup can be (partially) cancelled by
the inclusion of a term of the form\foot{In a previous
version of this paper, a different approach was employed which only
worked in the specific situation considered there.}
\eqn\counter{S'' =  {i \over 4\pi} \int d^{10} x \ln g(\tau) 
F \wedge X_8(R)\,,}
where $g(\tau)$ satisfies
\eqn\gcondone{ g(A\tau) = U_A
\left( {c\tau + d \over c \bar\tau + d} \right)^{1/2} g(\tau) \,,}
$A$ is the $SL(2,\IZ)$ transformation that is defined as in 
\modtrans, and $U_A$ is a constant phase. Under an $SL(2,\IZ)$
transformation the term $S''$ therefore transforms as
\eqn\trans{\delta_A S'' = - \delta_A S' 
+ i \ln U_A {1 \over 4\pi} \int d^{10}x\, F \wedge X_8(R) \,.}
The anomaly would be cancelled completely if it were possible to 
have  $U_A=1$ for all 
$A\in SL(2,\IZ)$, but as we shall demonstrate momentarily, this is not
possible. As a consequence only the $\tau$-dependent term in the
anomaly is removed by this term, and the condition that
$SL(2,\IZ)$ is unbroken leads to a non-trivial constraint on allowed
compactifications. 

The group of $SL(2,\IZ)$ is generated by the two transformations 
$T:\tau\mapsto\tau+1$ and $S:\tau\mapsto-1/\tau$ subject to the
relations $S^2=1=(TS)^3$. In terms of these generators, the condition
\gcondone\ becomes
\eqn\gcond{\eqalign{ g(\tau+1) & = U_T \, g(\tau) \cr
g(-1/\tau) & = U_S \, \left({\tau\over\bar\tau }\right)^{1/2} \,
g (\tau)\,,}}
where $U_S$ and $U_T$ are constant phases. Because of the square root
in \gcond\ there is a sign ambiguity, and we therefore only have the
relations $U_S^4=1=(U_T U_S)^6$. One solution to these conditions is
given by  
\eqn\gzero{
g_0(\tau) = {\eta(\tau) \over \eta(\tau)^*}  \,, }
where $\eta$ is the Dedekind eta-function
\eqn\etadef{\eta(\tau) = e^{i \pi\tau / 12} \prod_{n=1}^{\infty} 
( 1 - e^{2\pi i n \tau}) \,.}
In this case 
\eqn\compone{U^0_S=e^{i\pi/2} \qquad U^0_T=e^{i\pi/6}}
since $\eta$ transforms under the generators of $SL(2,\IZ)$ as 
\eqn\gendef{
\eta(\tau+1)  = e^{i\pi/12} \eta(\tau) 
\qquad
\eta(-1/\tau) = (-i \tau)^{1/2} \eta(\tau) \,.}
The most general solution of \gcond\ can then be written as 
$g(\tau) = g_0(\tau) \, h(\tau)$, where $h$ satisfies 
\eqn\gcond{\eqalign{ h(\tau+1) & = \hat{U}_T \, h(\tau) \cr
h(-1/\tau) & = \hat{U}_S \, h(\tau) \,.}}
Because of the relations of $SL(2,\IZ)$, $\hat{U}_S^2=1$ and 
$(\hat{U}_T \hat{U}_S)^3=1$ (where there is now no sign ambiguity),
and the possible values of $U_S$ and $U_T$ are 
\eqn\compgen{U_S = \hat{U}_S e^{i\pi/2} = \pm e^{i\pi/2} \qquad
U_T = \hat{U}_T e^{i \pi/6}\,.}
In particular, $U_S$ is always a (primitive) fourth root of unity, and  
$U_T$ is either a twelfth root of unity or a fourth root of unity. 
The former case is realised by the example \gzero\ above, and an
example of the latter case is given by 
\eqn\gone{ g_1(\tau) = g_0 (\tau) \left( 
{J(\tau)^* \over J(\tau)} \right)^{1/12} \,,}
where $J$ is the standard modular invariant $J$-function, and we have 
\eqn\jfunc{\eqalign{\left({J^*(\tau+1) \over J(\tau+1)} \right)^{1/12} 
& = e^{2\pi i /6} \left({J^*(\tau) \over J(\tau)} \right)^{1/12} \cr
\left({J^*(-1/\tau) \over J(-1/\tau)}\right)^{1/12}
& = - \left({J^*(\tau) \over J(\tau)} \right)^{1/12}\,.}}
In this case $U_S$ and $U_T$ generate a $\IZ_4$ subgroup of $U(1)$.

It is now clear that the $SL(2,\IZ)$ anomaly can only be cancelled up
to a term of the form
\eqn\constraint{ {2\pi \over N} {1\over 4\pi} 
\int d^{10}x\, F \wedge X_8 \,,}
where $N=4$ if $g$ has the same transformation properties as
$g_1$, and $N=12$ otherwise. In order for the $SL(2,\IZ)$ symmetry to
be unbroken, \constraint\ must be a multiple of $2\pi$, and hence the
compactification can only preserve the $SL(2,\IZ)$ symmetry if 
\eqn\finalcon{
{1 \over 4\pi} \int d^{10}x\, F \wedge X_8 \in 4 \IZ \,.}
This condition guarantees, in particular, that \consist\ is satisfied,
and therefore that the counterterm $S'$ is well-defined.

These considerations do not uniquely
determine the term $S''$. In particular, a given $g$ can always
be multiplied by a modular invariant function $\hat{h}$ without
modifying the anomaly cancellation. In principle it should be
possible to determine the actual form of the term $S''$
from a direct string calculation, but we shall not do so here.
However, if one were to assume that $g$ should be a product of a
holomorphic and anti holomorphic function and that $\ln g$ has a
perturbative term that is precisely $i\pi C^{(0)}/2$ then $g$ is
uniquely determined to be $g_1$ defined in \gone.

The anomaly can also be understood in the gauge fixed formulation: if
we fix the gauge by setting $\phi=0$, the counterterm $S'$ vanishes
but the fermion determinant transforms non-trivially under $SL(2,\IZ)$
transformations as follows from \compensate. This is again compensated
by the anomalous transformation of $S''$ up to the $\tau$-independent
terms discussed above.

\subsec{F-theory on K3}

In order to exhibit the constraint \finalcon\ more explicitly, let us 
consider the simplest example in which it leads to a non-trivial
condition. This is the theory where type IIB is compactified on the
euclidean product space $S^2 \times M^8$ that arises by inserting 24
seven-branes with locations at points in $S^2$. The world-volumes of
the branes are wrapped around the euclidean eight-manifold $M^8$ (we
are here envisioning a Wick rotation of the signature of both the
target space and the eight-volumes of the seven-branes). In terms of
F-theory this is the compactification on $K3\times M^8$. In order for
this compactification to preserve supersymmetry, $M^8$ has to be a
spin-manifold with a nowhere vanishing chiral spinor. 

In this case, using the fact that ${1\over 4\pi} \int F = 1$
\refs{\greene}, the consistency condition \finalcon\ simply becomes
\eqn\consistone{{1\over 4}\int d^{8} x \,X_8(R) 
= {\chi \over 24} \in \IZ\,,}
where $\chi$ is the Euler characteristic of the $M^8$ and 
we have used the relation
\eqn\nospin{\hat\chi = {1\over 8} (4p_2 - p_1^2)\,,}
which  is true for any eight-manifold  with a nowhere vanishing spinor
\refs{\popea} (and  $\hat\chi$ is the density which integrates to the
Euler character, $\chi=(2\pi)^{-4} \int \hat\chi$).
The identity \consistone\ also enters in discussions of the
compactification of type IIA superstrings \refs{\sensb}, $M$ theory
\refs{\beckera} and $F$ theory \refs{\vafwittb} on Calabi--Yau
four-folds (CY4). 

Th consistency condition \consist\ under which the $U(1)$ counterterm
$S'$ is well-defined is in this case
\eqn\consiste{ \int d^{8}x X_8(R) = {\chi \over 6} \in \IZ\,.}
It was shown in \refs{\vafwittb} that for the case of a Calabi-Yau
4-fold, $\chi/6$ is always an integer and therefore the consistency
condition \consist\ is satisfied.  It is also true that $\chi/6$ is an
integer for all of the currently known $Spin(7)$ manifolds listed in 
\refs{\joyce,\shatashvili}. It would be of interest if  this could be
proven to be true for all manifolds with a nowhere vanishing spinor.

\newsec{ Tadpole analysis}
 
It is instructive to contrast the above calculation with a tadpole
analysis of the F-theory compactification. The simplest examples of  
such compactifications \refs{\vafa} can be described by the insertion
of 24 parallel seven-branes in the type IIB vacuum.  Of these, up to
18 may be $(1,0)$ D7-branes while the rest must be other $(p,q)$ 
seven-branes.\foot{The coprime integers $p$ and $q$ label the \NSNS\
and \RR\ charges of the D-strings that can end on the various
seven-branes.}

Each of the 24 seven-branes in the F-theory
compactification has a term in its world-volume action of the form 
\refs{\bvs,\greenc,\dasguptaa}
 \eqn\worldvolfour{
- {1 \over 48} {1\over (2\pi)^2} \int C^{(4)} p_1\,,}
where $C^{(4)}$ is the (self-dual) RR
4-form of type IIB, and $p_1$ is the first Pontryagin class. For every
4-cycle $X_4$ in $M^8$ for which $\int_{X_4} p_1\ne 0$, type IIB
theory has a $C^{(4)}$ tadpole on the compact submanifold $X_4$, and
in order to cancel it, we have to include 
\eqn\dthree{
N(X_4)= 24 {1\over 48} {1\over (2\pi)^2} \int_{X_4} p_1}
D3-branes in the vacuum that are points on $X_4$ and extend along the
transverse directions. Since $X_8$ is a spin manifold, its second
Stiefel-Whitney class vanishes, and therefore $p_1/2$ is an integer
cohomology class of $X_8$;\foot{The reduction of the first Pontryagin
class $p_1$ mod 2 is precisely the square of the second
Stiefel-Whitney $w_2$ class on $X_8$, and thus vanishes for spin
manifolds \refs{\milnor}.} this implies that $N(X_4)$ is an integer for
every $X_4$. 

We now wish to determine the $C^{(0)}$ tadpole of the resulting
configuration.  Without loss of generality we may take sixteen of the
D7-branes to be $(1,0)$ seven-branes.  Each of these  has a term
\refs{\greenc} 
\eqn\wordvolzero{{1\over (2\pi)^4} {1\over (24)^2 \,40} 
\int C^{(0)} \, (9 p_1^2 - 8 p_2)\,, }
in its worldvolume action. Since $C^{(0)}$ is not invariant under
$SL(2,\IZ)$, it is not immediately clear what the contribution of the 
remaining eight seven-branes is. However, following Sen \refs{\sensb}
we can think of these eight seven-branes as corresponding to four
orientifold seven-planes that have each split into two seven-branes.  
At least at the point in moduli space where these two seven-branes are
on top of each other, we can describe them in terms of orientifold
planes, and we can thus summarize their gravitational effects by a
world-volume action of the form
\refs{\dasguptab}
\eqn\worldvolplane{{1\over (2\pi)^4} {1\over (24)^2 \, 20} 
\int C^{(0)} \, (27 p_1^2 - 44 p_2)\,.} 
Taking the contributions of the  sixteen D7-branes \wordvolzero\ and
the four orientifold seven-planes
\worldvolplane\ together, the total contribution of the 24
seven-branes to the $C^{(0)}$ tadpole 
becomes
\eqn\sevtot{\eqalign{
I_7 & = {1\over (2\pi)^4} 
\int_{M^8}  C^{(0)} \Bigl( {1\over 192} (p_1^2 - 4 p_2)
+ {1\over 96} p_1^2  \Bigr) \cr
  & = - {\chi \over 24} \, C^{(0)} 
  + {1\over (2\pi)^4}  \int_{M^8} {p_1^2 \over 96} \, C^{(0)} \,,\cr}}
where we have again used the  identity $\hat\chi = 8(4 p_2 - p_1^2)$. 
We also have to include the $C^{(0)}$ tadpole contribution of
the D3-branes, each of which has the term 
\eqn\fourzero{
- {1\over (2\pi)^2} {1 \over 48} \int C^{(0)} p_1\,,}
in their world-volume action. For every 4-cycle $X_4$, there are 
$N(X_4)$ D3-branes that extend along the transverse directions, and
the total contribution of the D3-branes to the $C^{(0)}$ tadpole is
therefore 
\eqn\tothr{\eqalign{I_3 & = -\sum_{X_4} N(X_4) 
{1\over (2\pi)^2} {1\over 48} 
\int_{X_4^\perp} p_1 C^{(0)}  \cr
& = - 24 {1\over (2\pi)^4} {1\over (48)^2} \int_{M^8} p_1^2\, C^{(0)}  
  = - {1\over 96} {1\over (2\pi)^4}
\int_{M^8} p_1^2\, C^{(0)} \,. \cr}}
Thus taking $I_7$ and $I_3$ together, the total $C^{(0)}$ tadpole is 
$-\chi/24 \; C^{(0)}$, which can be cancelled by the inclusion of 
$\chi/24$ D-instantons, provided that $\chi$ is again divisible by
$24$.  

As explained in \refs{\vafwittb}, the configuration breaks
supersymmetry unless the D-branes that are added to the background in
order to cancel the tadpoles are branes rather than anti-branes. This
leads to a further restriction on the signs of \dthree\ and $\chi$. 
It is difficult to establish these signs from first
principles, but we can determine them by considering special
examples.  For example, the
compactification on $K3\times K3$ preserves supersymmetry, and since 
${1\over (2\pi)^2} \int_{K3} p_1 = - 48$ and $\chi_{K3\times K3} =
24^2$, we deduce that $\chi$ has to be a positive integer and $N(X_4)$
a negative integer. 

The above analysis applies to an arbitrary 8-dimensional manifold that
has a covariantly constant spinor, including Calabi--Yau four-folds
and $Spin(7)$ manifolds. The $Spin(7)$ manifolds constructed by Joyce
\refs{\joyce} all have Euler character $\chi =6\cdot 24$ (and
signature $\sigma=64$), and thus satisfy the above condition.   
Additional $Spin(7)$ manifolds have been found for which the signature
is different from $64$, but the Euler character of all known examples
is divisible by $24$ \refs{\shatashvili}. These examples all arise as
blow-ups of (symmetric) geometrical 
toroidal orbifolds, and one may expect that
the corresponding string theory (or $F$-theory) is necessarily
consistent. This would `explain' why the Euler character of all of
these examples  is divisible by $24$. This explanation should also
apply in the case of compactifications on  Calabi--Yau four-folds,
which generically have Euler characters that are  multiples of $6$
only \refs{\vafwittb}. There again, the consistency of string
compactifications should imply that any four-fold that is  constructed
by blowing up a symmetric orbifold should have an Euler number that
is a multiple of $24$.

\newsec{Relation to the heterotic string}

The compactification of F theory  on (elliptically fibred) $K_3$ is
supposed to be equivalent to the heterotic string compactified on
$T^2$ \refs{\vafa}, and thus the same F-theory features that we have
discussed should be apparent from the point of view of the heterotic
string.  The eight-dimensional theory has a duality symmetry group
which contains $SL(2,\IZ) \times SL(2,\IZ)$, where the two $SL(2,\IZ)$
are associated with transformations of the K\"ahler class and the
complex structure of the two-torus, $T^2$. For comparison with the
F-theory discussion it is sufficient to consider generic Wilson lines
on $T^2$ which break the heterotic gauge symmetry to $U(1)^{20}$. The
eight-dimensional low energy action, which is  the Einstein--Maxwell
action of \refs{\salama,\townsends} with eighteen vector multiplets,
has chiral couplings that give rise to an anomaly in a  $U(1)$ current
that is embedded in an $SL(2,\IR)$ subgroup of the classical symmetry
group of the theory. This  $U(1)$ anomaly of the eight-dimensional
supergravity theory is essentially  the same  anomaly that we
discussed earlier in a somewhat different F-theory guise. As before,  
the anomaly can be expressed  as a local $U(1)$ anomaly
together with an anomaly cancelling term that does not preserve
$SL(2,\IZ)$. In this case the local anomaly has the form,
\eqn\hetanom{ \int d^8x \left(X_8 + 12
I_{1/2}\right)\, \Sigma(x)\,}
and the anomaly cancelling term has a form analogous to
\anomcancel. This term again induces an $SL(2,\IZ)$ anomaly that can
be partially cancelled by a term analogous to \counter. When the
theory is compactified on an eight-manifold $M^8$, the whole of 
$SL(2,\IZ)$ is unbroken provided that $\int X_8/4 = \chi/24$ is again
an integer; since $\int 3I_{1/2}$ is always integer, the second term
in \hetanom\ does not affect this result. 

Another problem with the eight-dimensional heterotic string, arises
from the well-known term that is associated with the cancellation of
ten-dimensional gravitational (and gauge) anomalies. After
compactification on $T^2$ this term has the form   
$\int d^8x\, B Y_8(R)$, where $B\equiv B_{9\,10}$ and
\refs{\greenschwarza} 
\eqn\anomdef{ Y_8 ={1\over 4}  X_8 + 15 I_{1/2} = 
{1\over 128 (2\pi)^4} \left(\tr R^4 +
{1\over 4} (\tr R^2)^2\right)\,.}
Upon further  compactification  on $M^8$ this gives rise to a $B$
tadpole which is the same as the $C^{(0)}$ tadpole arising from the
compactification of the 24 F-theory seven-branes in the last section.
This tadpole is cancelled by wrapping the world-volumes of $N(X_4)$
five-branes around the torus and the transverse four-cycles in
$M^8$, and by wrapping the world-sheets of $\chi/24$ fundamental
strings around $T^2$. Again this is only possible if $\chi/24$ is a
positive integer.  
 
\vskip 0.8cm
To summarize, we have seen that type IIB supergravity has a local
$U(1)$ anomaly that can be cancelled by a local counterterm which
induces an anomaly in the global  $SL(2,\IR)$ transformations; this
means that $SL(2,\IR)$ is not a symmetry of the theory.  
However, the $\tau$-dependence of the anomaly in an $SL(2,\IZ)$
subgroup is cancelled by the inclusion of another term 
($S^{\prime\prime}$) and the remaining $\tau$-independent 
anomaly is irrelevant in the path-integral provided that the
background satisfies a certain topological restriction. For F-theory
compactifications of the form $K3\times M^8$, the requirement is that
the Euler character of the eight-dimensional manifold is a multiple of
$24$. This condition is compatible with (although slightly weaker
than) the condition that arises from the requirement of tadpole
cancellation in F-theory compactifications. Similar considerations
also apply to the dual heterotic string theory.

\vskip 0.3cm
\noindent{\it Acknowledgments}\hfill\break
We are grateful to Ashoke Sen, Terry Gannon and Savdeep Sethi for
useful comments.  We also wish to thank Constantin Bachas and
Pascal Bain for illuminating discussions that led to the revisions in
this version of the paper.

M.R.G. is grateful to Jesus College, Cambridge for a Research
Fellowship, and to Fitzwilliam College, Cambridge, for a College
Lectureship. This Research was also supported in part by PPARC.

\listrefs
\end